\documentclass[twocolumn,showpacs,preprintnumbers,amsmath,amssymb,superscriptaddress]{revtex4}

\newtheorem{theorem}{Theorem}
\newtheorem{proposition}{Proposition}

\begin{document}

\title{Infinite number of conditions for local mixed state manipulations}  
\author{Gilad Gour}\email{ggour@math.ucsd.edu}
\affiliation{Department of Mathematics, University of California/San Diego, 
        La Jolla, California 92093-0112}

\date{\today}

\begin{abstract} 
It is shown that a finite number of conditions are {\em not}
sufficient to determine the locality of transformations between two
probability distributions of pure states as well as the locality of
transformations between two $d\times d$ mixed states with $d\geq 4$. 
As an example, an infinite, but minimal, set of necessary and
sufficient conditions  for the existence of a local procedure that
converts one probability distribution of two pure pair of qubits into
another one is found. 
\end{abstract}


\maketitle

Entanglement is one of the main ingredients of non-intuitive
quantum phenomena. Besides of being of interest from a fundamental point of 
view, entanglement has been identified as a non-local resource for quantum 
information processing. In particular, shared bipartite entanglement is a crucial 
resource for many quantum information tasks~\cite{NC} and therefore
its quantification is very important. Pure bipartite
entanglement
is quantified {\em asymptotically} by the entropy of
entanglement~\cite{BBPS}
and {\em deterministically}  by a set of entanglement
monotones~\cite{Vidal,Vidal2}. 
Nevertheless, {\em mixed} state entanglement is far more rich and lacks 
a complete quantification despite the enormous efforts that has 
been made in the last years~\cite{Horodecki}. Since in practice one
usually works with mixed states, quantification of mixed state
entanglement is extremely important for quantum information 
processing.

In this paper we show a surprising result on the number of
measures of entanglement that are required to quantify the non-local
resources of bipartite mixed states. An entangled pair of
qudits (each of dimension $d$) is described by a finite dimensional
density matrix $\rho$ with $d^{4}$ elements. Therefore, a finite number
of independent measures of entanglement (i.e. entanglement monotones) 
are required to quantify completely the non-local resources of $\rho$;
that is, there must be a finite set of entanglement monotones such that
{\em any} measure of entanglement can be expressed as a function
of these monotones~\footnote{For example, the set of entanglement
monotones given in Eq.(\ref{one}) (see~\cite{Vidal2}), or the set 
of concurrence monotones given 
in~\cite{Gour04}, are both quantifying completely the non-local resources 
of a pure bipartite state, in the sense that any measure of 
entanglement can be expressed in terms of these entanglement monotones.} 
Yet, it is shown below that for $d\geq 4$ a finite number of 
measures of entanglement is {\em insufficient} to determine whether a 
transformation $T:\;\rho\rightarrow\sigma$ can be realized by means of 
local operations and classical communication (LOCC). That is, for {\em any}
finite number of measures of entanglement, say $\{E_k\}$, there exist
$\rho$ and $\sigma$ such that $E_{k}(\rho)\geq E_{k}(\sigma)$ (for all
$k$) and yet $\rho$ can not be transformed to $\sigma$ by LOCC. 
As will be shown in the following, similar results  
hold also for converting one probability distribution
of {\em pure} states to another.    

For a given transformation, $T$, between two finite dimensional bipartite pure
entangled states, there is a set of entanglement monotones that provide
necessary and sufficient conditions to determine if $T$ is 
local (i.e. can be realized by LOCC).
The family of entanglement monotones $E_{k}$ ($k=0,1,2,...,d-1$)
which introduced in~\cite{Vidal2} 
were first defined over the set of pure states as
\begin{equation}
E_{k}(|\psi\rangle)=\sum_{i=k}^{d-1}\lambda _{i}\;,
\label{one}
\end{equation}
where $\lambda_{0}\geq\lambda_{1}\geq\cdots\geq\lambda_{d-1}$ are the 
Schmidt numbers of the $d\times d$-dimensional bipartite
state $|\psi\rangle$, and then extended to mixed states by means of the convex roof 
extension. For a pure state $|\psi\rangle$ these measures of entanglement quantify 
{\em completely} the non-local resources since all the Schmidt coefficients of 
$|\psi\rangle$ are determined by them. According to Nielsen
theorem~\cite{Nielsen}
the transformation between two bipartite states
$T:\;|\psi\rangle\rightarrow |\phi\rangle$ can be performed by LOCC
iff
\begin{equation}
E_{k}(|\psi\rangle)\geq E_{k}(|\phi\rangle)\;\;\;\forall\;k\;.
\label{two}
\end{equation}
This result has been extended by Jonathan and Plenio~\cite{Jon99} to
the case where the transformation $T$ lead to several possible
final states. That is, 
\begin{equation}
T:\;|\psi\rangle\rightarrow D\;,
\label{33} 
\end{equation}
where 
$D=\{p_{i},|\phi_{i}\rangle\}$ is a probability distribution of final 
states with $p_i$ being the probability that $T$ outputs the state 
$|\phi _{i}\rangle$. It has been shown in~\cite{Jon99} that
T is local iff
\begin{equation}
E_{k}(|\psi\rangle)\geq E_{k}(D)\equiv 
\sum_{i}p_{i}E_{k}(|\phi\rangle _{i})\;\;\;\forall\;k\;.
\label{three}
\end{equation}

There are other sets of entanglement monotones, such as the concurrence 
monotones~\cite{Gour04}, that quantify completely the non-local resources 
of a pure state. However, the advantage of the entanglement monotones 
defined in Eq.(\ref{one}) is that they provide {\em sufficient} 
conditions for the transformation $T$ in Eq.(\ref{33}) to be local.
In the following we show that a finite set of entanglement monotones with 
this property does not exist (1) if $T$ represents a
transformation between two probability distributions of pure bipartite states,
$T:\;D_{1}\rightarrow D_{2}$, and (2) if $T$ represents a transformation 
between two mixed bipartite states $T:\;\rho\rightarrow\sigma$.

In order to prove the main result of this paper (theorem 1)
we first consider a transformation between two probability distributions.
Suppose Alice and Bob share a pure state $|\psi\rangle$ and then perform 
quantum operations (not necessarily local) represented by a 
transformation, $t$, that outputs the state $|\psi_{i}\rangle$ 
($i=1,2,...,n_1$) with probability $p_i$. We then say that Alice and Bob
share a probability distribution denoted by 
$D_{1}=\{p_i,\;|\psi_{i}\rangle\}_{i=1}^{n_1}$. Suppose now that for each 
$i$, if the transformation $t$ outputs the state
$|\psi_{i}\rangle$, Alice and Bob perform again quantum operations which 
are represented by the transformation $T_{i}$. We denote by $\{|\phi_j\rangle\}$ 
($j=1,2,...,n_2$) all the possible outcome states such that each 
transformation $T_i$ outputs the state $|\phi _{j}\rangle$ with 
conditional probability $q_{j|i}$. Thus, the transformation 
$T\equiv (T_1,T_2,...,T_{n_{1}})$  
between the probability distributions $D_1$ and 
$D_2\equiv\{q_j,\;|\phi_j\rangle\}_{j=1}^{n_2}$, outputs the state
$|\phi_j\rangle$ with probability $q_{j}=\sum_{i}p_{i}q_{j|i}$. We start 
with the most simple, but non-trivial, case in which $n_1=2$ and both 
$|\psi_1\rangle$ and $|\psi_2\rangle$ are $2\times2$-dimensional pure 
states. We also take $n_2=1$; that is, the distribution $D_2=|\phi\rangle$, 
where $|\phi\rangle$ is also a $2\times2$-dimensional pure state.   
 
For any entangled pair of qubits with Schmidt numbers $\lambda _{0}$ and $\lambda _{1}$
(i.e. $|\psi\rangle=\sqrt{\lambda _{0}}|00\rangle+\sqrt{\lambda _{1}}|11\rangle$)
we define the parameter $x=2{\rm min}\{\lambda _{0},\lambda _{1}\}$. Note that $0\leq x\leq 1$,
and from Eq.(\ref{one}) $x=2E_{1}(|\psi\rangle)$; i.e. $x$ measure the entanglement of 
$|\psi\rangle$. Moreover, any measure of entanglement, $E$, for a pure pair of qubits
can be written as a function of $x$; that is, $E(|\psi\rangle)=f(x)$. The function $f$
is not necessarily continuous (for example, take $E(|\psi\rangle)$ to measure the Schmidt 
number of $|\psi\rangle$), but it must satisfy the following conditions:\\
(i) Monotonicity: for $x_1 \geq x_2$, $f(x_1)\geq f(x_2)$\\
(ii) Concavity: $f(\sum p_{i}x_{i})\geq\sum p_{i}f(x_{i})$\\
(iii) Normalization: $f(0)=0$ and $f(1)=1$\\

Conditions (i) and (ii) are necessary for entanglement
monotones~\cite{Vidal,Nielsen,Jon99} whereas condition (iii) is useful in comparison
between different measures of entanglement so that for all measures of
entanglement the Bell state has one unit of entanglement. Moreover,
these three conditions are sufficient to prove the following simple two
lemmas, which will be useful for later:

{\it Lemma 1}: $f(x)\geq x$

{\it Proof}: $f(x)=f\left(x\cdot 1+(1-x)\cdot 0\right)$ and from 
  conditions (ii) and (iii) we have
$f(x)\geq xf(1)+(1-x)f(0)=x$.

{\it Lemma 2}: If $f(x_1)=f(x_2)$ for $x_1\neq x_2$ then
$f(x_1)=f(x_2)=1$.

{\it Proof}: We assume that $x_1 < x_2 < 1$; thus, there is $0<t<1$ 
such that $x_2=tx_1+1-t$. From condition (ii) we have 
$f(x_2)\geq tf(x_1)+1-t$, and since $f(x_1)=f(x_2)$ we get 
$f(x_1)\geq 1$. From (i) and (iii) we have $f(x_1)\leq 1$ and 
therefore $f(x_1)=f(x_2)=1$.

We are now ready to prove the following observation (proposition 1). 
For simplicity
we will restrict
our attention to probability distributions $D_1,\;D_2$ of initial and
final $2\times 2$ dimensional states with $n_1=2$ and $n_2=1$. The
proposition
is trivially applicable for higher
dimensions and/or for higher $n_1$ and $n_2$. 
\begin{proposition}
For {\em any} finite number of measures of 
entanglement $\{E_k\}_{k=1}^{s}$ ($s<\infty$), there are probability
distributions $D_1$ (with $n_1=2$) and $D_2$ (with $n_2=1$) such that
\begin{equation}
E_k(D_1)\geq E_k(D_2)\;\;\forall\;\; k=1,2,...,s
\label{theorem1}
\end{equation} 
although the transformation $T:\;D_1\rightarrow D_2$ can not be
realized by LOCC with certainty.
\end{proposition}

{\bf Proof}: Let us denote by $x_1,\;x_2$ and $y$ twice of the minimal
Schmidt numbers of $|\psi_1\rangle,\;|\psi_2\rangle$ and
$|\phi\rangle$,
respectively. With these notations, Eq.~(\ref{theorem1}) can be 
rewritten as
\begin{equation}
p_1f_k(x_1)+p_2f_k(x_2)\geq f_k(y)\;\;\forall\;\; k=1,2,...,s\;,
\label{t1}
\end{equation}
where $\{f_k\}_{k=1}^{s}$ is a finite set of functions that 
satisfy conditions (i)-(iii) above. Thus, we would like first 
to find $x_1,\;x_2,\;y$ and $p_1$ ($p_2=1-p_1$) such that 
Eq.(\ref{t1}) is satisfied for all $k$.
Now, the set $\{f_k\}$ may
include the Schmidt function (i.e. $f(x=0)=0$ and $f(x>0)=1$).
Besides the Schmidt function, for all the other functions there exist
$x>0$ such that $f_k(x)<1$. Therefore, since the set $\{f_k\}$ is
finite, there exist $y>0$ such that besides the Schmidt function,
$f_k(y)<1$ for all $k$. We can also find $0<p_1<1$ such that
$f_k(y)\leq p_1$ for all $k$ (besides the Schmidt function). Thus,
by taking $x_1=1$, the inequality in Eq.(\ref{t1}) is satisfied for 
{\em any} value of $0<x_2\leq 1$. Note that even for the Schmidt function
the inequality is satisfied. However, the transformation 
$T:\;D_1\rightarrow D_2$ can not be
realized by LOCC if $x_2<y$. That is, the state $|\psi_2\rangle$
occurs
with probability $p_2=1-p_1>0$ and according to Nielsen theorem the
transformation $|\psi_2\rangle\rightarrow|\phi\rangle$ can be realized
by LOCC if and only if $y\geq x_2$. Thus, by taking $0<x_2<y$ we prove
proposition 1 $\Box$

In the above observation we see that finite number of conditions are 
insufficient to determine if a transformation between two probability 
distributions, $T:\;D_1\rightarrow D_2$, can be realized locally. A natural 
question that presents itself is then: what are the sufficient conditions for 
$T$ to be local? In the general case, the answer to this question 
appears to be complicated especially since it involves
infinite number of conditions. Nevertheless, in the following 
proposition we provide sufficient conditions for the case in which both $D_1$ and
$D_2$ denotes probability distributions of two $2\times 2$ dimensional
pure states. 

\begin{proposition}
Let two distant parties share a probability distribution, $D_1$, of 2
(initial) $2\times 2$-dimensional pure states $|\psi_1\rangle$ and
$|\psi_2\rangle$ with corresponding probabilities $p_1$ and $p_2$, 
respectively. Let also $D_2$ be a probability distribution of 2 (final) 
$2\times 2$-dimensional pure states $|\phi_1\rangle$ and 
$|\phi_2\rangle$ with corresponding probabilities $q_1$ and $q_2$,
respectively. Then the transformation $T:\;D_1\rightarrow D_2$ can be
realized by LOCC if, and only if,
\begin{equation}
E_{\mu}(D_1)\geq E_{\mu}(D_2)\;\;\;\forall\;0\leq\mu\leq 1\;,
\label{theorem2}
\end{equation}
where
\begin{align}
E_{\mu}(|\psi\rangle)\equiv f_{\mu}(x)\equiv
\left\{
\begin{array}{ll} 
& x/\mu\;\;{\rm for}\;\;x\leq\mu\\
& \;1\;\;\;{\rm for}\;\;x>\mu\\
\end{array}\right.
\end{align}
\end{proposition}
Note that $f_{\mu=0}$ is the Schmidt function and $f_{\mu=1}(x)=x$.  

{\bf Proof}: It is easy to see that the functions $f_{\mu}$ satisfy
conditions (i)-(iii); that is, they are entanglement monotones. As
such, the inequalities in Eq.(\ref{theorem2}) are {\em necessary} 
conditions that any local transformation $T$ must satisfy. We will now
show that they are also {\em sufficient}.

Denoting by $x_1$ and $x_2$ twice the minimal Schmidt number of 
$|\psi_1\rangle$ and $|\psi_2\rangle$, respectively, we
first note that according to Eq.~(\ref{three}) if $x_1=x_2$ the condition 
$E_{\mu=1}(D_1)\geq E_{\mu=1}(D_2)$ is a sufficient condition for $T$
to be local~\cite{Jon99}. Thus, without loss of generality
we take $x_1>x_2$. We also denote by $y_1$ and $y_2$ twice the minimal 
Schmidt number of $|\phi_1\rangle$ and $|\phi_2\rangle$, respectively, 
and assume that $y_1\geq y_2$. Now, if
$y_2>x_2$ we get $E_{\mu=y_2}(D_1)< E_{\mu=y_2}(D_2)=1$; that is, the
condition in Eq.(\ref{theorem2}) is not satisfied for $\mu=y_2$. Thus we
have $x_2\geq y_2$. 

The most general transformation $T:\;D_1\rightarrow D_2$
consist of two transformations which we denote by $T_1$ and $T_2$.
The transformation $T_1$ ($T_2$) on $|\psi_1\rangle$ ($|\psi_2\rangle$)
outputs the states $|\phi_1\rangle$, $|\phi_2\rangle$ with
conditional probabilities $q_{1|1}$, $q_{2|1}=1-q_{1|1}$ 
($q_{1|2}$, $q_{2|2}=1-q_{1|2}$), respectively. Thus, the
probabilities $q_{1|1}$ and $q_{1|2}$ must satisfy 
$q_1=p_1q_{1|1}+p_2q_{1|2}$ (or equivalently
$q_2=p_1q_{2|1}+p_2q_{2|2}$). According to Eq.~(\ref{three}), $T_1$ and
$T_2$ can be realized by LOCC (i.e. $T$ can be realized by LOCC) if,
and only if,
\begin{align}
& x_1\geq q_{1|1}y_1+(1-q_{1|1})y_2\nonumber\\
& x_2\geq q_{1|2}y_1+(1-q_{1|2})y_2\;.\label{cond}
\end{align}

We now consider the three possible options:\\ 
(a) $x_1>x_2\geq y_1\geq y_2$: in this case we take
$q_{1|1}=q_{1|2}=q_1$. According to Eq.~(\ref{cond}) the transformations
$T_1$ and $T_2$ can be realized by LOCC. Hence,
the transformation $T=(T_1,T_2):\;D_1\rightarrow D_2$ can also be
realized by LOCC.\\
(b) $x_1\geq y_1>x_2\geq y_2$: in this case, according to
Eq.~(\ref{cond}) the transformation $T_1$ 
can be realized by LOCC for any value of $q_{1|1}$. On the other hand, the
transformation $T_2$ can be realized by LOCC only if 
$q_{1|2}\leq q_{1|2}^{max}\equiv (x_2-y_2)/(y_1-y_2)$. Thus, the transformation 
$T=(T_1,T_2):\;D_1\rightarrow D_2$ can be
realized by LOCC only if $p_1+q_{1|2}^{max}p_2\geq q_1$. By substituting the
value of $q_{1|2}^{max}$, this condition can be expressed as:
\begin{equation}
p_1y_1+p_2x_2\geq q_1y_1+q_2y_2\;.
\end{equation}
Now, it can be shown that this condition is satisfied by taking
$\mu=y_1$ in~Eq.~(\ref{theorem2}).\\ 
(c) $y_1\geq x_1>x_2\geq y_2$: in this case, according to
Eq.~(\ref{cond}), the transformations $T_1$ and $T_2$ can be realized
by LOCC only if $q_{1|1}\leq q_{1|1}^{max}\equiv (x_2-y_2)/(y_1-y_2)$ and 
$q_{1|2}\leq
q_{1|2}^{max}$. Thus, the transformation 
$T$ can be realized by LOCC only if $q_{1|1}^{max}p_1+q_{1|2}^{max}p_2\geq q_1$. 
By substituting the values of $q_{1|1}^{max}$ and $q_{1|2}^{max}$, this condition
can be expressed as: 
\begin{equation}
p_1x_1+p_2x_2\geq q_1y_1+q_2y_2\;.
\end{equation} 
This condition is obtained by taking $\mu=1$
in~Eq.~(\ref{theorem2}) $\Box$

In the observation above the set of conditions given in Eq.~(\ref{theorem2}) is 
{\em not} minimal. In fact, it is enough to require that
$E_\mu(D_1)\geq E_\mu(D_2)$ for all the {\em rational} numbers $\mu$ in the
interval $[0,1]$. The proof for proposition 2 will not change much because
for any irrational number $\mu\in [0,1]$, there is a series of rational
numbers $\{\mu_n\}_{n=1}^{\infty}$ ($\mu_n\in [0,1]$) such that 
$\lim _{n\rightarrow\infty}\mu _n=\mu$. According to proposition 1 there
must be an infinite number of conditions and therefore the set of
conditions in Eq.(\ref{theorem1}) with rational $\mu$ is 
countable and in this sense is minimal (but not unique). We are now ready to prove 
the main result of this paper:  

\begin{theorem}
Let $\{E_k\}_{k=1}^{s}$ be a finite set of entanglement monotones
defined on $d\times d$-dimensional bipartite mixed states. Then, for
$d\geq 4$ there exist density matrices $\rho$ and $\sigma$ such that:\\
(i) $E_k(\rho)\geq E_k(\sigma)\;\;\forall\;\;k=1,2,...,s$\\
(ii) There is no local procedure that converts $\rho$ into $\sigma$
(i.e. the transformation $T:\;\rho\rightarrow\sigma$ can not be
realized by LOCC with certainty).
\end{theorem}

{\bf Proof}: The density matrix $\rho$ is taken to be of the following
form: 
\begin{equation}
\rho=p_1|\psi_1\rangle\langle\psi_1| +p_2|\psi_2\rangle\langle\psi_2|\;,
\end{equation}
where $p_1$ and $p_2$ will be determined later, and
\begin{align}
&
|\psi_1\rangle=\frac{1}{\sqrt{2}}\left(|00\rangle+|11\rangle\right)\nonumber\\ 
&
|\psi_2\rangle=\sqrt{\lambda}|22\rangle+\sqrt{1-\lambda}|33\rangle\;, 
\end{align}
where the Schmidt number $\lambda$ is smaller then 1/2 and
will be determined later. The density matrix $\sigma$ is taken to be
the pure state:
\begin{equation}
\sigma=|\phi\rangle=\sqrt{\eta}|00\rangle+\sqrt{1-\eta}|11\rangle\;,
\end{equation}
with Schmidt number $\eta < 1/2$.

Suppose now that Alice performs a projective measurement with
projectors $P_0\equiv |0\rangle\langle 0|+|1\rangle\langle 1|$ and 
$P_1\equiv |2\rangle\langle 2|+|3\rangle\langle 3|$. The result of her
measurement yields the state $|\psi_1\rangle$ with probability $p_1$
and the state $|\psi_2\rangle$ with probability $p_2$. Thus, any
measure of entanglement must satisfy $E(\rho)\geq
p_1E(|\psi_1\rangle)+p_2E(|\psi_2\rangle)$. On the other hand, 
$E(\rho)\leq
p_1E(|\psi_1\rangle)+p_2E(|\psi_2\rangle)$ because if
Alice and Bob forget the result of the measurement they end up 
back with the same state $\rho$. Thus, any measure of entanglement, $E$, 
must satisfies  
\begin{equation}
E(\rho) = p_1E(|\psi_1\rangle)+p_2E(|\psi_2\rangle)\;.
\end{equation}

Now, following the same lines of proposition 1, there exist values of
$p_1$, $\lambda$ and $\eta$ such that (1) $E_k(\rho)\geq E_k(|\phi\rangle)$
for all $k$ and (2) $\eta >\lambda >0$. We would like to show now that
Alice and Bob can not convert $\rho$ to $\phi$ by LOCC. For this
purpose, we define a set of entanglement monotones similar to the one
defined in Eq.~(\ref{theorem2}). 

For any given normalized measure of
entanglement~\footnote{The measure of entanglement, $E$, is normalized such 
that $E=1$ for a maximally entangled state.}, $E$, we define a set of 
entanglement monotones as
follows  
\begin{align}
E_{\mu}(|\psi\rangle)\equiv
\left\{
\begin{array}{ll} 
& \mu^{-1}E(|\psi\rangle)\;\;{\rm if}\;\;E(|\psi\rangle)\leq\mu\\
& \;1\;\;\;{\rm if}\;\;E(|\psi\rangle)>\mu\\
\end{array}\right.
\label{mm}
\end{align}
where $\mu\in [0,1]$ and the definition of $E$ and $E_\mu$ for mixed
state are given in terms of the convex roof extension. Since we assume that
$E$ is an entanglement monotone, it follows from Theorem 2
in~\cite{Vidal} that the $\{E_\mu\}$ are indeed entanglement monotones for all
$0\leq\mu\leq 1$. Let us now take the measure of entanglement, $E$,
given in~Eq.~(\ref{mm}) to be
\begin{equation}
E(|\psi\rangle)=\frac{4}{3}(1-\lambda_{max})\;,
\label{mm1}
\end{equation}
where $\lambda_{max}$ is the largest Schmidt number of $|\psi\rangle$.
Thus, for $\mu=\eta$, $E_\mu$ as defined in Eqs.~(\ref{mm},\ref{mm1}) 
satisfies
\begin{equation}
E_{\mu=\eta}(\rho)=\frac{2}{3}p_1+\frac{4}{3}(1-\lambda)p_2
<\frac{2}{3}=E_{\mu=\eta}(|\phi\rangle)\;.
\end{equation}
That is, there exist entanglement monotone that quantify
$|\phi\rangle$ with more entanglement than $\rho$. Thus, Alice and Bob
can not convert $\rho$ into $\sigma$ by LOCC $\Box$

Any finite dimensional density matrix, $\rho$, consists of a finite
number of parameters, $\rho_{ij}$. This means that any measure of
entanglement is a function of these parameters. Thus, there must be a
finite number of entanglement monotones, say $\{E_k\}_{k=1}^{s}$, that
quantify {\em completely} the non-local resources of
$\rho$. That is, all measures of entanglement can be written as a
function of these $s$ entanglement monotones. Nevertheless, according
to theorem 3, these measures of entanglement are not sufficient to
determine whether $\rho$ can be converted into $\sigma$ by LOCC. A
natural question is then arise: is there an {\em infinite} number of
entanglement monotones that do provide
the sufficient conditions for a transformation $T:\;\rho\rightarrow\sigma$
to be local? We generalize this question to include probability
distributions of mixed states in the following conjecture: 

{\em Conjecture}: A transformation $T:\;D_1\rightarrow D_2$ between
two probability distributions of bipartite mixed states
can be realized by LOCC iff
$E(D_1)\geq E(D_2)$ for {\em all} entanglement monotones $E$.

The necessity of these conditions is trivial to prove though the 
sufficiency appears to be complicated to prove due to the 
complexity of mixed state entanglement. The conjecture above has been  
proved for the case where $D_1$ is a single pure bipartite 
state~\cite{Nielsen,Jon99} and in proposition 2 we have proved 
this conjecture for the case where $D_1$ and $D_2$ each is
a probability distribution of two {\em pure} pair of qubits. 
If the conjecture above is incorrect for the general case, then
entanglement would be insufficient to quantify non-locality in quantum
mechanics. Since we have some examples with the flavor of
``quantum non-locality without entanglement''~\cite{Ben99}, it 
would not be too surprising (though very interesting) if the conjecture 
above turns out to be incorrect. 

To summarize: we have shown that infinite number of conditions (based
on entanglement monotones) are required for determining the locality
of transformations on a given $d\times d$-dimensional mixed bipartite
state (with $d\geq 4$) and on a given probability distribution of pure
bipartite states. We have also presented a minimal set of infinite 
number of conditions that are required for determining the locality of
transformations between two probability distributions of two 
pure pair of qubits. We believe that our results will also prove fruitful
in further developments on mixed state entanglement. 

\textbf{Acknowledgments:} I am indebted to David Meyer for very valuable 
discussions and to Joe Henson for his help. This work has been partially 
supported by the Defense Advanced Research Projects Agency (DARPA) QuIST 
program under contract F49620-02-C-0010, and by the National Science
Foundation (NSF) under grant ECS-0202087.

\end{document}